\documentclass[namedreferences]{solarphysics}

\usepackage[hyperref]{spr-sola-addons} 
\usepackage{graphicx}        
\usepackage{color}           
\usepackage{breakurl}        

\usepackage{caption}
\usepackage{soul}
\usepackage{array, multirow}

\chardef\us=`\_

\begin{document}

\begin{article}
\begin{opening}

\title{The behaviour of galactic cosmic ray intensity during solar activity cycle 24\\ {\it Solar Physics}}

\author[addressref={aff1,aff2},corref,email={exr007@student.bham.ac.uk}]{\inits{E.}\fnm{Eddie}~\lnm{Ross}\orcid{0000-0003-4437-7910}}
\author[addressref={aff1,aff2},corref,email={w.j.chaplin@bham.ac.uk}]{\inits{W.J.}\fnm{William J.}~\lnm{Chaplin}}


\address[id=aff1]{School of Physics and Astronomy, University of Birmingham, Edgbaston, Birmingham B15 2TT, UK}
\address[id=aff2]{Stellar Astrophysics Centre (SAC), Department of Physics and Astronomy, Aarhus University, Ny Munkegade, DK-8000 Aarhus C, Denmark}

\runningauthor{Ross et al.}
\runningtitle{GCR behaviour during solar cycle 24}

\begin{abstract}
We have studied long-term variations of galactic cosmic ray (GCR) intensity in relation to the sunspot number (SSN) during the most recent solar cycles. This study analyses the time-lag between the GCR intensity and SSN, and hysteresis plots of the GCR count rate against SSN for solar activity cycles 20-23 to validate a methodology against previous results in the literature, before applying the method to provide a timely update on the behaviour of cycle 24. Cross-plots of SSN vs GCR show a clear difference between the odd-numbered and even-numbered cycles. Linear and elliptical models have been fit to the data with the linear fit and elliptical model proving the more suitable model for even-numbered and odd-numbered solar activity cycles respectively, in agreement with previous literature. Through the application of these methods for the 24th solar activity cycle, it has been shown that cycle 24 experienced a lag of 2-4 months and follows the trend of the preceding activity cycles albeit with a slightly longer lag than previous even-numbered cycles. It has been shown through the hysteresis analysis that the linear fit is a better representative model for cycle 24, as the ellipse model doesn't show a significant improvement, which is also in agreement with previous even-numbered cycles.
\end{abstract}
\keywords{Cosmic Rays, Galactic; Solar Cycle, Observations; Sunspots, Statistics}
\end{opening}

\section{Introduction}
     \label{S-Introduction} 
     
Galactic cosmic rays (GCRs) are charged particles and atomic nuclei with energies spanning the range from a few MeV up to approximately $10^{21}$ eV, that encroach upon the Earth from all directions  \citep{giacalone_energetic_2010}. They mainly originate outside the solar system, within the Milky Way; however they are also expected to originate from other galaxies \citep{aab_observation_2017}. GCRs at the top of the atmosphere are mostly composed of protons ($\sim87$\%) and $\alpha$-particles ($\sim12$\%), with a smaller contribution ($\sim1$\%) from heavier nuclei \citep{dunai_cosmic_2010}.

When cosmic rays (CRs) enter the atmosphere, they interact with atmospheric atoms and produce cascades of secondary particles, which at ground level are primarily neutrons and muons. Neutron monitors (NMs) and muon detectors (MDs) located at different locations on Earth have been used since the 1950s to observe GCRs. Information on GCRs prior to the modern epoch of NMs and MDs, and the space age, rely on the studies of cosmogenic isotope records from ice cores and tree rings \citep{owens_heliospheric_2013}.

\begin{figure}
	\includegraphics[width=\columnwidth]{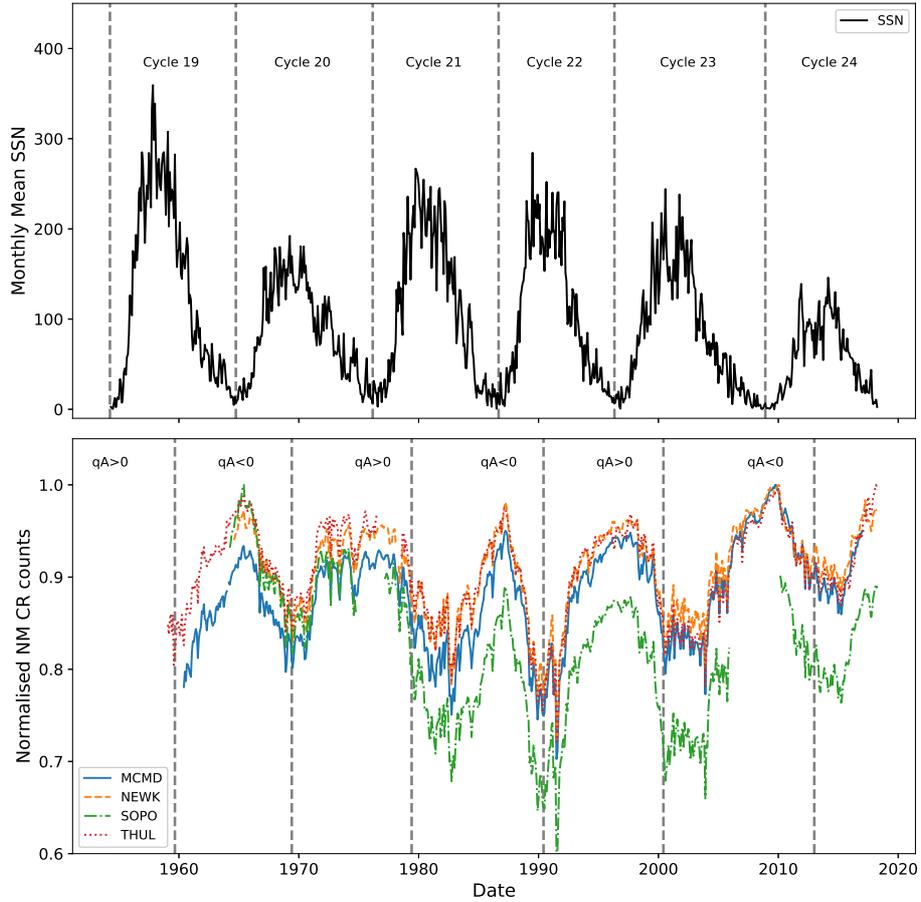}
    \caption{SSN (top), with vertical lines showing the beginning of each solar cycle. Cosmic ray intensity recorded by NMs (bottom), with vertical lines showing the approximate epochs of solar magnetic field polarity reversals.
    (MCMD = McMurdo, NEWK = Newark, SOPO = South Pole, THUL = Thule).}
    \label{fig:CR_SSN_timeseries}
\end{figure}

It has long been established that there exists an anti-correlation between GCR intensity and the level of solar activity, over a cyclic 11-year period, with perhaps some time-lag  \citep{forbush_cosmic-ray_1958, parker_passage_1965, usoskin_correlative_1998, van_allen_modulation_2000}. Figure \ref{fig:CR_SSN_timeseries} shows clearly the anti-correlation between GCRs and sunspot number (SSN).

It is well known that the 11-year solar activity cycle is in fact a 22-year cycle -  the Hale cycle - which describes the alternating polarity of the large-scale solar magnetic field  \citep{thomas_22-year_2014}. The interchanging peaked and flat-topped shape of GCR intensity in Figure \ref{fig:CR_SSN_timeseries} is a manifestation of this effect in addition to other CR transport processes \citep{aslam_solar_2012}.

The polarity of the solar field, $A$, is taken to be negative when the field axis is aligned with the axis of rotation, and positive when the opposite is true \citep{thomas_22-year_2014}. The solar field polarity conventionally is described in combination with particle charge, $q$, due to the effect of curvature and gradient drift on charged particles; thus it is customary to define the solar polarity as $qA$. Vertical lines showing the approximate epochs at which the polarity reverses are plotted in Figure \ref{fig:CR_SSN_timeseries} \citep{janardhan_solar_2018, thomas_22-year_2014}. 

Particle drifts differ during different $qA$ cycles, with positive CRs (i.e. protons) predominantly arriving into the heliosphere from the heliospheric poles and outwards to Earth during periods when $qA > 0$, whereas when $qA< 0$, positive CRs predominantly arrive at Earth inwards along the heliospheric current sheet (HCS) \citep{belov_large_2000, thomas_galactic_2014}. As the solar magnetic dipole axis is tilted to the solar rotation axis, so is the HCS; the HCS tilt varies with solar cycle and is typically smaller during solar minimum and larger during solar maximum \citep{owens_heliospheric_2013}. The tilt angle of the HCS has also been shown to be strongly correlated to the GCR intensity and the GCR lag behind the solar activity \citep{belov_large_2000, mavromichalaki_cosmic-ray_2007}. 

\cite{aslam_solar_2012, aslam_study_2015} found that the different processes of CR transport have varying levels of importance throughout the solar activity cycle, but around solar maximum it is likely that drifts play less of a role and disturbances in the solar wind (and hence HCS) are the predominant factor of CR modulation. Even cycles encounter $qA<0$ polarity during their onset phase and $qA>0$ during their declining phase, thus experiencing a faster GCR recovery after solar maximum as the GCRs predominantly enter the heliosphere from the heliospheric poles and experience an outwards drift towards Earth. Odd cycles encounter $qA>0$ polarity during their onset phase and $qA<0$ during their declining phase and so experience a slower recovery after solar maximum, as the GCRs predominantly enter the heliosphere along the HCS. When the HCS is tilted and disturbed during the declining activity phase, the path length that GCRs must travel to Earth increases; hence resulting in an increased time-lag.

Several studies have demonstrated the lag between GCR and solar activity proxies is approximately zero (i.e. no lag) during even solar cycles, and that there exists a lag of around a year or more during odd solar cycles \citep{usoskin_correlative_1998, mavromichalaki_cosmic-ray_2007, singh_solar_2008}.

\begin{figure}[ht]
	\includegraphics[width=\columnwidth]{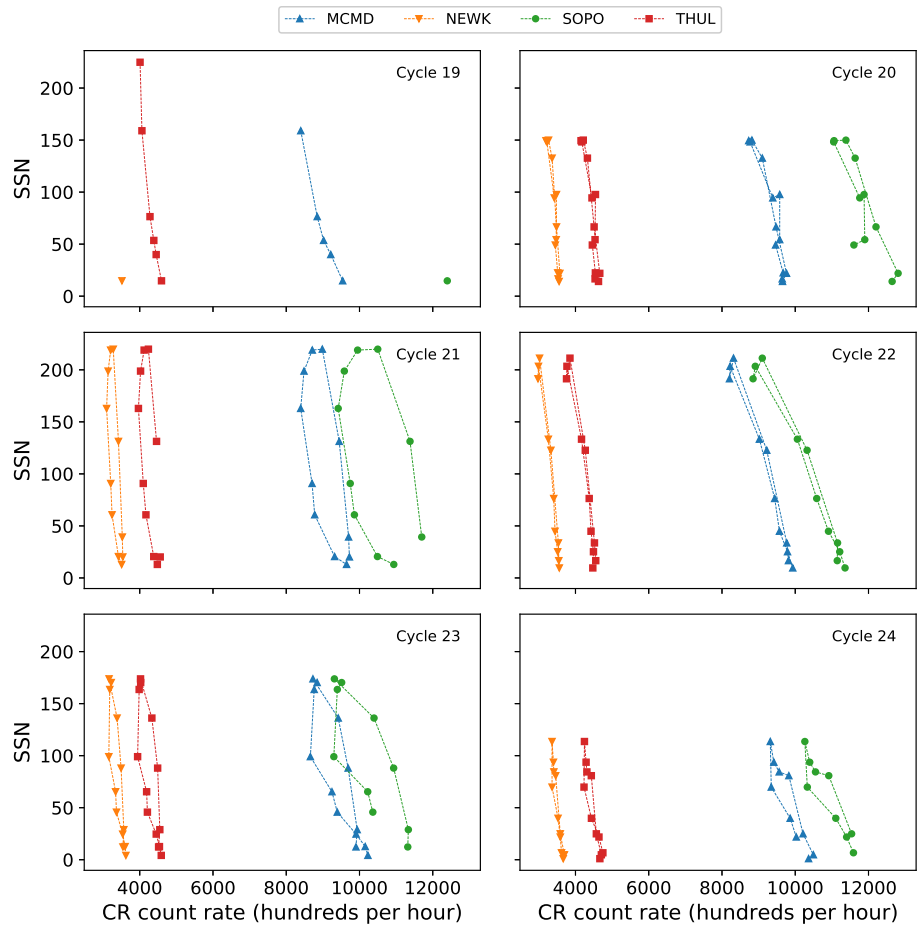}
    \caption{Hysteresis plots between yearly averaged SSN and yearly averaged GCR intensity for each of the 4 main NM stations over cycles 19-24.}
    \label{fig:all_hysteresis}
\end{figure}

\citet{van_allen_modulation_2000} showed through cross-plotting the annual mean intensity of GCRs against sunspot number (SSN) between 1953 and 1999 (covering solar cycles 19-22), that there is a distinct difference in the plot shapes between the different solar cycles, with 19 and 21 producing broad ovals, and 20 and 22 as approximately flat lines. The striking difference between odd-numbered and even-numbered cycles is shown for cycles 19-24 in Figure \ref{fig:all_hysteresis}. It is believed that this hysteresis effect is caused by the combination of the heliospheric magnetic field (HMF), solar magnetic field polarity and thus the particle drift, and the tilt of the HCS leading to a slow recovery of GCR intensity after maxima in odd cycles and a fast recovery after maxima in even cycles  \citep{van_allen_modulation_2000, belov_large_2000, thomas_22-year_2014}.

An extension of this work has since been carried out by \citet{inceoglu_modeling_2014} showing that the even numbered solar activity cycles can be best modeled using a linear fit due to the narrow shape of the hysteresis loops; whereas odd-numbered solar activity cycles are better represented by ellipses due to their broader shape.

There has been speculation in the literature on the behaviour of cycle 24 compared to recent odd and even cycles. It has been suggested that there exists a lag between SSN maxima and GCR intensity minima in excess of 10 months \citep{kane_lags_2014, mishra_study_2016} which does not follow the previous even cycles having a near-zero lag and in fact suggesting that cycle 24 behaved similarly to previous odd cycles; however these studies do not make use of a complete cycle of data and thus may draw inaccurate conclusions about the behaviour of the whole cycle because of the unusually extended nature of the declining phase of cycle 23 and the low amplitude of cycle 24 maximum \citep{broomhall_helioseismic_2017}. \cite{mishra_study_2016} make use of a more complete data set for cycle 24, yet still incomplete, and conclude that it is also likely that a 4 month lag could exist between GCRs and SSN.

This work aims to provide a timely update on the statistical relationship between GCR intensity and solar activity during solar cycle 24, since the cycle has now almost declined to a minimum. These aims have been achieved through a time-lag analysis and hysteresis effect analysis between SSN and GCR intensity.

In Section \ref{S-Data} we provide a brief description of the data that was used throughout this study for both CRs and SSN.

We show in Section \ref{sec:lag} through a correlative time-lag analysis that there exists a small time-lag between the SSN and GCR intensity over solar cycle 24, which is slightly longer than preceding even-numbered cycles but not as high as observed in previous odd-numbered cycles. We also discuss whether the time-lag between SSN and GCR shows a dependence on the rigidity cut-off of the observing station. 

In Section \ref{sec:hysteresis} we model the shapes of hysteresis plots between GCR intensity and SSN. We show that the behaviour of the hysteresis loops for cycle 24 follow the preceding even-numbered solar activity cycles and is better represented by a straight line fit rather than an elliptical model.

\section{Data}
     \label{S-Data} 

For the majority of the work in this study we have considered the pressure corrected count rates measured by four NM monitor stations as acquired from the NM data base \citep{nmdb_nmdb_nodate} event search tool (NEST) (\url{http://nmdb.eu/nest/}). The four stations are McMurdo (MCMD), Newark (NEWK), South Pole (SOPO), and Thule (THUL), i.e. the same NM stations used in the study by \citet{inceoglu_modeling_2014} to provide a comparison to existing literature. Table \ref{tab:NM_stns} details the basic characteristics of the NM stations used in this study. 

We have investigated the long-term GCR modulation in the heliosphere from 1964-2018, spanning solar cycles 20-24, for the cycle epochs: 20: (10/1964 - 03/1976); 21: (03/1976 - 09/1986); 22: (09/1986 - 05/1996); 23: (05/1996 - 12/2008); 24: (12/2008 - 03/2018). Early predictions on solar cycle 25 suggest that solar cycle 24 is unlikely to reach a minimum earlier than the middle of 2019 up to as far as early 2021 (see \cite{howe_signatures_2018, upton_updated_2018, pesnell_early_2018}). The data used in this study are therefore of an incomplete cycle 24; however we believe this to now have a minimal effect on the results as cycle 24 draws to a minimum. Cycle 19 was omitted from this study due to the incomplete data set for this period (see Figure \ref{fig:CR_SSN_timeseries} and Figure \ref{fig:all_hysteresis}).

\begin{table}

 \caption{Neutron monitor stations used in this study and their vertical geomagnetic cut-off rigidity (R$_c$), longitude, latitude, and altitude acquired from NEST. The first four stations have been used for all of the analysis while the lower 12 stations have been used exclusively for the investigation into the dependence of $R_c$ on the time-lag.}
 \label{tab:NM_stns}
 
 \begin{tabular}{l l l c c c c}
  \hline
  		& &  & $R_c$  & Long. & Lat.  & $h$  \\
  		& & Station & [GV] & [deg] & [deg] & [m]  \\
  \hline
\parbox[t]{2mm}{\multirow{4}{*}{\rotatebox[origin=c]{90}{\tiny Time-Lag \& }}} &
\parbox[t]{2mm}{\multirow{4}{*}{\rotatebox[origin=c]{90}{\tiny Hysteresis }}}
		 & McMurdo (MCMD) & 0.30 & 166.6 E & 77.9 S & 48 \\
		& & Newark (NEWK) & 2.40 & 75.8 W & 39.7 N & 50  \\
		& & South Pole (SOPO) & 0.10 & 0.0 E & 90.0 S & 2820\\
		& & Thule (THUL) & 0.30 & 68.7 W & 76.5 N & 26 \\
		  \hline
\parbox[t]{2mm}{\multirow{12}{*}{\rotatebox[origin=c]{90}{\tiny R$_c$ dependence}}} &
\parbox[t]{2mm}{\multirow{12}{*}{\rotatebox[origin=c]{90}{\tiny of time-lag }}}
		 & Oulu (OULU) & 0.81 & 25.5 E & 65.1 N & 15 \\
		& & Kerguelen (KERG) & 1.14 & 70.3 E & 49.4 S & 33 \\
		& & Magadan (MGDN) & 2.10 & 151.1 E & 60.0 N & 220 \\
		& & Climax (CLMX) & 3.00 & 106.2 W & 39.4 N & 3400  \\
		& & Dourbes (DRBS) & 3.18 & 4.6 E & 50.1 N & 225 \\
		& & IGY Jungfraujoch (JUNG) & 4.49 & 7.98 E & 46.6 N & 3570 \\
		& & Hermanus (HRMS) & 4.58 & 19.2 E & 34.4 S & 26 \\
		& & Alma-Ata B (AATB) & 6.69 & 76.9 E & 43.0 N &  3340 \\
		& & Potchefstroom (PTFM) & 6.98 &  27.1 E &  26.7 S &  1351 \\
		& & Mexico (MXCO) & 8.28 & 99.2 W & 19.8 N & 2274 \\
		& & Tsumeb (TSMB) & 9.15 & 17.6 E & 19.2 S &  1240 \\
		& & Huancayo (HUAN) & 12.92 &  75.3 W &  12.0 S &  3400 \\
  \hline
  
 \end{tabular}
 
\end{table}


During the time-lag correlation analysis, as our results suggested there may be a rigidity dependence on the time lag, we introduced a further 12 NM stations with data acquired from NEST spanning cycles 20-24 to increase the rigidity spectrum utilised in this study; these stations and their basic characteristics are also detailed in Table \ref{tab:NM_stns}. These stations are not included in the rest of the results however as the results from these stations do not change the conclusions of this study.

Furthermore, we have also used monthly/yearly averaged SSN, as collected by WDC-SILSO (\url{http://sidc.be/silso/}), for the time-lag analysis/hysteresis analysis respectively as our chosen proxy of solar activity.

\section{Time-Lag Analysis}
\label{sec:lag}

\subsection{Method}
\label{TL-Method} 
To investigate the time delay between the modulation of GCRs compared to the solar activity, a time-lag cross-correlation analysis was performed between monthly mean GCR intensity and monthly mean SSN for each station, following the approach of \citet{usoskin_correlative_1998}. We used a time window of width $T$ centred on a time $t$, i.e. shifting within the interval $t-T/2$ to $t+T/2$. Here we used $T = 50$ months.

The window was shifted in steps $\Delta t = 1$ month within this interval and for each step the Spearman's rank correlation coefficient ($\rho$) between GCR intensity and SSN was calculated. The lag between GCR and SSN was then estimated by finding the peak correlation coefficient within the time interval $T$.

The results from the four main NM stations used suggested that there may be a relationship between the rigidity cut-off ($R_c$) of a NM station and the time-lag for GCRs; hence four additional NM stations were introduced to determine whether this was so, as detailed above.


\subsection{Results}

\begin{figure}[ht!]
	\includegraphics[width=\columnwidth]{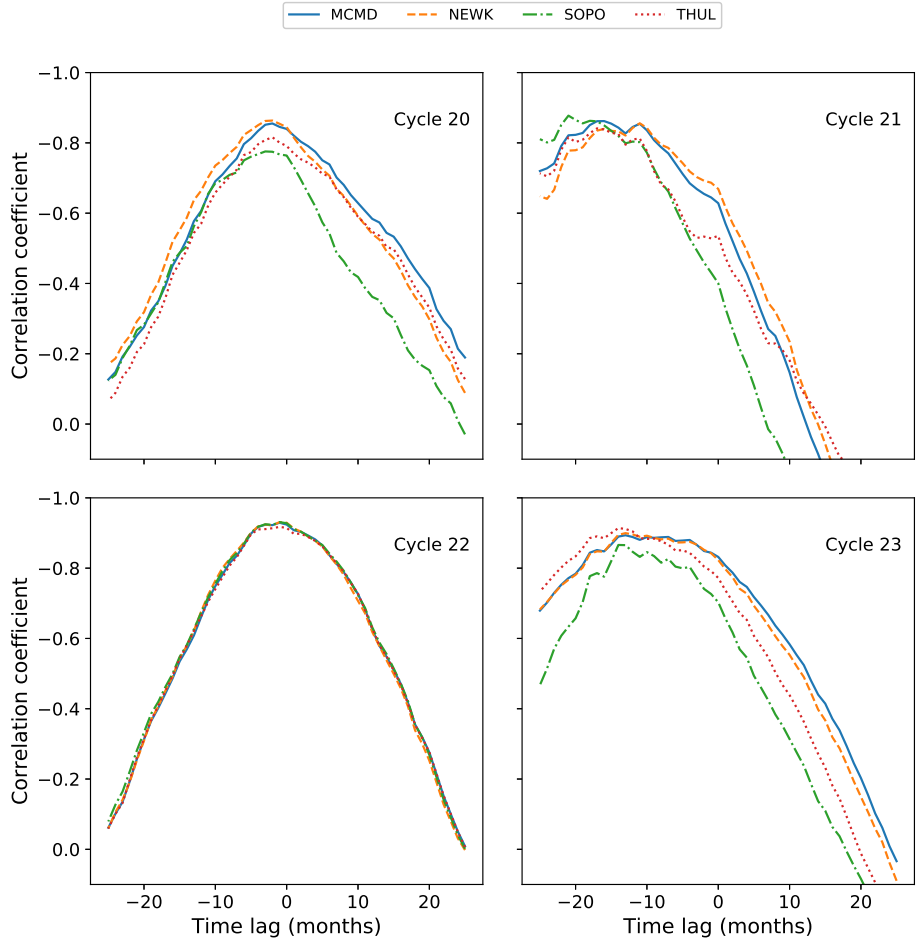}
    \caption{Variation in the correlation coefficient with time-lag NM station GCR intensity and SSN during solar cycles 20-23.}
    \label{fig:time_lags_20-23}
\end{figure}

The correlation ($\rho$) between monthly averaged GCR counts and SSN for different time-lags was calculated for cycles 20-23. The variation in $\rho$ is presented in Figure \ref{fig:time_lags_20-23}, showing that for each cycle there is a time-lag corresponding to peak anti-correlation between GCR intensity and SSN. Table \ref{table:time_lags_20-23} summarises the time-lag with the highest correlation and the corresponding correlation coefficient for all stations in each individual solar cycle.

\begin{table}[!ht]
\caption{Time-lags and the corresponding cross-correlation coefficient between NM CR count and SSN for solar cycles 20-23.}
\label{table:time_lags_20-23}
\begin{tabular}{l c c c c c c c c}
	\hline 
	{} & \multicolumn{2}{c}{Cycle 20} & \multicolumn{2}{c}{Cycle 21}  \\
	{} & {Lag [months]} & {$\rho$} & {Lag [months]} & {$\rho$} \\ \hline
	{McMurdo} & {2} & {$-0.855$} & {16} & {$-0.862$}  \\
	{Newark} & {2} & {$-0.863$} & {11} & {$-0.856$} \\
	{South Pole} & {3} & {$-0.776$} & {21} & {$-0.877$}  \\
	{Thule} & {2} & {$-0.816$} & {17} & {$-0.841$} \\ \hline
	
	{} & \multicolumn{2}{c}{Cycle 22} & \multicolumn{2}{c}{Cycle 23}\\
	{} & {Lag [months]} & {$\rho$} & {Lag [months]} & {$\rho$}\\ \hline
	{McMurdo}  & {1} & {$-0.929$} & {13} & {$-0.894$} \\
	{Newark} & {1} & {$-0.931$} & {13} & {$-0.900$} \\
	{South Pole}  & {1} & {$-0.931$} & {14} & {$-0.866$} \\
	{Thule}  & {1} & {$-0.917$} & {14} & {$-0.914$}\\ \hline
\end{tabular}
\end{table}

As previously reported in the literature, we see here that all of the NM stations clearly exhibit almost no lag during even solar cycles, and a longer lag varying between 11-21 months during odd solar cycles. There is a strong agreement between the results presented in Table \ref{table:time_lags_20-23} and those of \cite{mavromichalaki_cosmic-ray_2007}, \cite{kane_lags_2014}, and \cite{paouris_solar_2015-1}, thus providing further evidence on the distinction between odd and even solar cycles due to particle transport in the heliosphere. The agreement with existing literature provides evidence of a suitable methodology in this study.

The same cross-correlation technique was then applied to cycle 24 between the dates 12/2008 - 03/2018 and the results are presented in Figure \ref{fig:time_lag_24} and Table \ref{table:time_lags_24}.

\medskip
%
%
{\centering
\begin{minipage}{0.52\textwidth}
   \centering
   \includegraphics[width=\columnwidth]{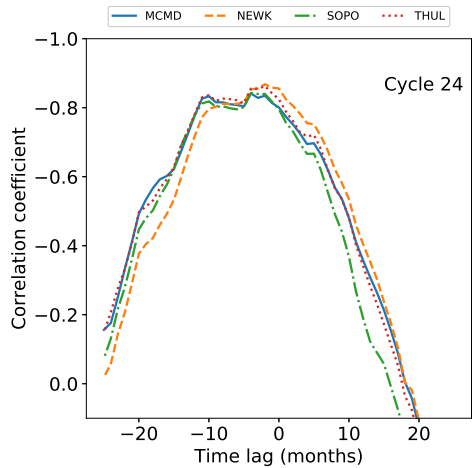}
   \captionsetup{font=scriptsize}
   \captionof{figure}{Variation in the correlation coefficient with time-lag between NM GCR intensity and SSN during solar cycle 24.}
   \label{fig:time_lag_24}
\end{minipage}
\hspace{0.05\textwidth}
\begin{minipage}{0.41\textwidth}
   \centering
   \captionsetup{font=scriptsize}
   \captionof{table}{Time-lags and the corresponding cross-correlation coefficient between NM GCR intensity and SSN for solar cycle 24.}
   \label{table:time_lags_24}
   \normalsize
   \begin{tabular}{l c c c c c c c c}
	\hline 
	{} & \multicolumn{2}{c}{Cycle 24} \\
	{} & {Lag} & {$\rho$} \\ 
	{} & {[months] } & {} \\ \hline
	{MCMD} & {4} & {$-0.841$} \\
	{NEWK} & {2} & {$-0.868$} \\
	{SOPO} & {4} & {$-0.843$} \\
	{THUL} & {2} & {$-0.862$}  \\ \hline
\end{tabular}
\end{minipage}
 \hfill \break }
 \medskip

Cycle 24 is seen here to follow the pattern of almost no lag for even cycles; however cycle 24 does display a lag that is larger than the previous two even-numbered cycles, despite not being as large as the two previous odd-numbered cycles. The cause for the increased time-lag in cycle 24, as compared to the previous two even-numbered cycles, is likely due to the combined effects of the unusually deep and extended minimum between solar cycles 23 and 24, which delayed the decline in GCR intensity and caused record-breaking high GCR intensities \citep{pacini_unusual_2015}, and the small amplitude of the cycle 24 maximum.

The results presented in this study, using data for a near-complete cycle 24, show that the results of \cite{kane_lags_2014}, and \cite{mishra_study_2016}, were likely unduly influenced by the unusually deep and extended declining phase of cycle 23 given that they had a limited data set. \cite{mishra_study_2016} used data for just over half of cycle 24 and resulted in a time-lag of 4 months which agree with the results of this study.

As a further note on time-lag, \cite{tomassetti_evidence_2017} showed that through the introduction of time-lag as a parameter in the CR transport calculations of CR spectra that there exists a time-lag of $8.1 \pm 1.2$ months during the period 2000-2012 spanning across cycle 23 and 24. We performed the time-lag analysis for the first 4 NM stations detailed in Table \ref{tab:NM_stns} for the period between 2000-2012 to investigate whether these results can be reproduced. The results of this analysis are presented in Figure \ref{fig:time_lag_tomasetti} and Table \ref{table:time_lags_tomasetti}.

\medskip
%
%
{\centering
\begin{minipage}{0.52\textwidth}
   \centering
   \includegraphics[width=\columnwidth]{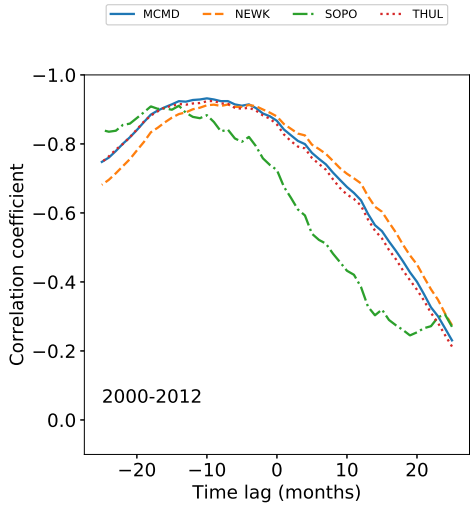}
      \captionsetup{font=scriptsize}
   \captionof{figure}{Variation in the correlation coefficient with time-lag between NM GCR intensity and SSN between 2000-2012.}
   \label{fig:time_lag_tomasetti}
\end{minipage}
\hspace{0.03\textwidth}
\begin{minipage}{0.42\textwidth}
   \centering
      \captionsetup{font=scriptsize}
   \captionof{table}{Time-lags and the corresponding cross-correlation coefficient between NM GCR intensity and SSN during 2000-2012. }
   \label{table:time_lags_tomasetti}
   \begin{tabular}{l c c c c c c c c}
	\hline 
	{} & \multicolumn{2}{c}{2000-2012} \\
	{} & {Lag} & {$\rho$} \\ 
	{} & {[months] } & {} \\ \hline	
	{ MCMD} & {10} & {$-0.932$} \\
	{ NEWK} & {7} & {$-0.914$} \\
	{ SOPO} & {14} & {$-0.911$} \\
	{ THUL} & {9} & {$-0.925$}  \\ \hline
\end{tabular}
\end{minipage}
 \hfill \break }
 \medskip

From the time-lag analysis of the 4 stations presented there is a mean lag of $10.00 \, \pm \, 1.47$ months, which is in good agreement with the results of \cite{tomassetti_evidence_2017}.

\begin{figure}
	\includegraphics[width=\columnwidth]{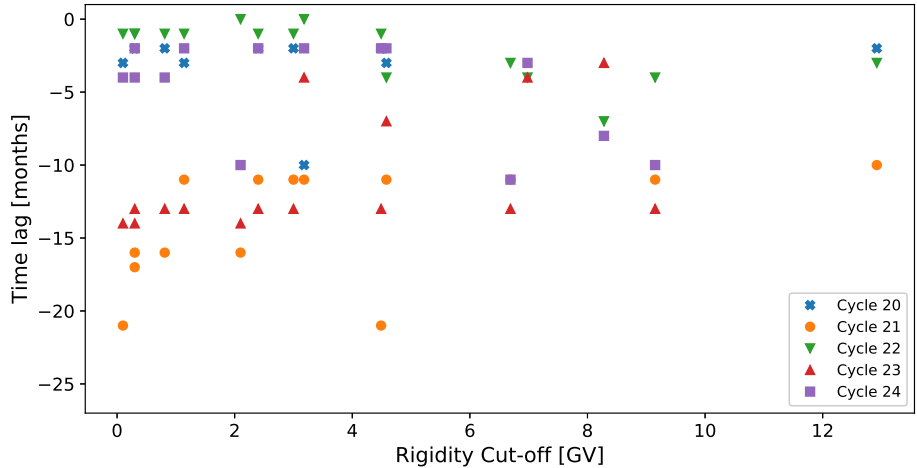}
    \caption{Variation in time-lag plotted against NM station rigidity cut-off for the 16 NM stations detailed in Table \ref{tab:NM_stns}.}
    \label{fig:time_lags_rigidity}
\end{figure}

Finally, allowing for the odd/even cycle dependence, we see in Figure \ref{fig:time_lags_20-23}, Figure \ref{fig:time_lag_24}, and Figure \ref{fig:time_lag_tomasetti} that the time-lag appears to depend on the rigidity of the NM station used for observation. Such a dependence may impact the conclusions depending on the choice of NM station. We expect that if a dependence exists, a station with a higher rigidity cut-off ($R_c$) would have a shorter lag as this station observes higher energy CRs which are affected less by solar modulation and thus able to recover faster from solar maximum. Whereas a station with a lower cut-off rigidity observing lower energy CRs, which are more influenced by solar modulation, would recover more slowly from solar modulation and therefore experience a longer time-lag. This is supported by Figure \ref{fig:time_lags_20-23}, Figure \ref{fig:time_lag_24}, and Figure \ref{fig:time_lag_tomasetti}, but in order to provide more conclusive evidence of such a relationship we introduced the additional NM stations detailed in Table \ref{tab:NM_stns} to provide a rigidity range spanning 0-13 GV. We present in Figure \ref{fig:time_lags_rigidity} a plot of the time-lag versus station $R_c$ for all 16 stations over cycles 20-24. To acquire uncertainties on the time-lag we ran 1000 Monte Carlo simulations of the time-lag analysis, sampling from the uncertainty distributions for each of the monthly averaged SSN and GCR counts; however the uncertainties in the data propagated in the Monte Carlo simulations produced no scatter in the overall results.

The results of this analysis do not show a clear rigidity dependence on the time-lag between SSN and GCR intensity; the sampling of higher $R_c$ is too low, due to the availability of high $R_c$ stations, to reasonably conclude on such a dependence at high rigidities, despite cycle 21 suggesting the existence of a dependence for low $R_c$ stations as per our expectations. For low $R_c$ stations there appears to be a more pronounced distinction between the time-lag observed between odd-numbered and even-numbered cycles than for higher $R_c$ stations, however again this is not definitive due to the low sampling at higher $R_c$. We therefore conclude that there will be no significant dependence of the time-lag analysis on the $R_c$ of the observing station.

\section{Hysteresis Effect Analysis}
\label{sec:hysteresis}

\subsection{Method}
\label{HA-Method} 

To investigate the hysteresis effect, we have adopted the approaches of \cite{van_allen_modulation_2000}, \cite{singh_solar_2008}, and \cite{inceoglu_modeling_2014}. Plots of the annual mean SSN versus the annual mean GCR intensity were generated for cycle 20-24 and analysed by fitting different models to the data.

As highlighted in \citet{inceoglu_modeling_2014}, even-numbered solar cycles can be suitably modeled by a linear fit due to their narrow hysteresis shape, and odd-numbered solar cycles were shown to be better modeled by ellipses due to their broadened hysteresis shape. Here, we first repeated for solar cycles 20-23 the linear and ellipse fitting to confirm the method reproduces the results reported in \citet{inceoglu_modeling_2014} before applying the method to cycle 24.

For even cycles which display narrow hysteresis loops, an unweighted least squares linear regression was used to reconstruct estimates of the GCR intensity from SSN. As odd-numbered solar cycles display a broader hysteresis loop, they were separately modeled using unweighted linear regression and ellipse fitting to determine the model providing the better fit. 

The equation of the ellipse fitting took the form:

\begin{equation}
	\left[ \begin{array}{c} x \\ y \end{array} 	\right] = 
	\left[ \begin{array}{c} x_0 \\ y_0 \end{array} \right] + 
	R(\phi)
	\left[ \begin{array}{c} a \, \cos{\theta} \\ b \, \sin{\theta} \end{array} \right]
	\label{eq:ellipse}
\end{equation}

where $x$ is GCR intensity; $y$ is SSN; $(x_0, y_0)$ are the centroid coordinates of the fitted ellipse, $R(\phi)$ is the rotation matrix; $\phi$ is the ellipse tilt angle; $a$ and $b$ are the semi-major and semi-minor axes respectively; and $0 \, \leq \, \theta \, \leq 2\pi$ is the polar angle measured anti-clockwise from the semi-major axis.

In order to regain the GCR intensity from the model, where linear regression was used to model the data, GCR was acquired directly from the SSN for each year. For the ellipse model the GCR was acquired from the model as a function of time from $\theta$, where the time-lag calculated from the analysis in Section \ref{sec:lag} was used to correctly phase the ellipse allowing $\theta$ to be calculated using standard equations of ellipses.

GCR intensity predicted by the linear regression and ellipse models were compared to the measured GCR intensity using Spearman's rank correlation as per \citet{inceoglu_modeling_2014}.

\subsection{Results}

The hysteresis loops between yearly averaged SSN and GCR intensity for each station were first modeled with a linear regression for both odd and even solar activity cycles, then the odd cycles were separately re-modeled by ellipse fitting to show that this provides a more representative fit as suggested in \cite{inceoglu_modeling_2014}. The correlation between measured CR intensities and modeled CR intensities for cycles 20-23 are presented in Table \ref{table:hysteresis_20-23}.

\begin{table}[!ht]
\caption{Correlation coefficients of the linear regression and ellipse modeling of the hysteresis plots for solar cycles 20-23.}
\label{table:hysteresis_20-23}
\begin{tabular}{l c c c c c c c c}
	\hline 
	{} & \multicolumn{2}{c}{Cycle 20} & \multicolumn{2}{c}{Cycle 21} & \multicolumn{2}{c}{Cycle 22} & \multicolumn{2}{c}{Cycle 23}\\
	{} & {Linear} & {Ellipse} & {Linear} & {Ellipse} & {Linear} & {Ellipse} & {Linear} & {Ellipse}\\ \hline
	{McMurdo} & {0.867} & {-} & {0.664} & {0.946} & {0.964} & {-} & {0.846} & {0.852} \\
	{Newark} & {0.888} & {-} & {0.700} & {0.964} & {0.955} & {-} & {0.857} & {0.874} \\
	{South Pole} & {0.746} & {-} & {0.358} & {0.939} & {0.936} & {-} & {0.733} & {0.855} \\
	{Thule} & {0.783} & {-} & {0.912} & {0.964}  & {0.900} & {-} & {0.813} & {0.929}\\ \hline
\end{tabular}
\end{table}

\begin{figure}
	\includegraphics[width=\columnwidth]{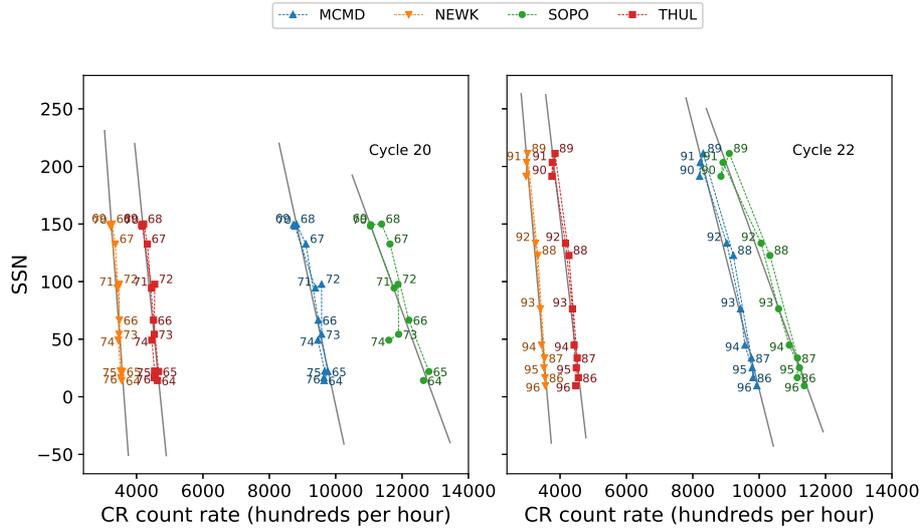}
    \caption{The hysteresis plots for even solar cycles 20 and 22 and the linear regression fit to the data.}
    \label{fig:linear_even}
\end{figure}

There is a consistent and good agreement between the measured and modeled CR intensities for even solar cycles modeled through linear regression, because the hysteresis loops are quite narrow as shown in Figure \ref{fig:linear_even}. These results support the findings of \cite{inceoglu_modeling_2014}. Note that discrepancies in the correlation coefficients between this study and \cite{inceoglu_modeling_2014} are likely due to a number of reasons: \cite{inceoglu_modeling_2014} used data smoothing processes where in this study raw data are used; \cite{inceoglu_modeling_2014} made use of monthly mean data, whereas annual mean data was used in this study; \cite{inceoglu_modeling_2014} interpolated missing data, whereas gaps have been left untreated in this study.

\begin{figure}
	\includegraphics[width=\columnwidth]{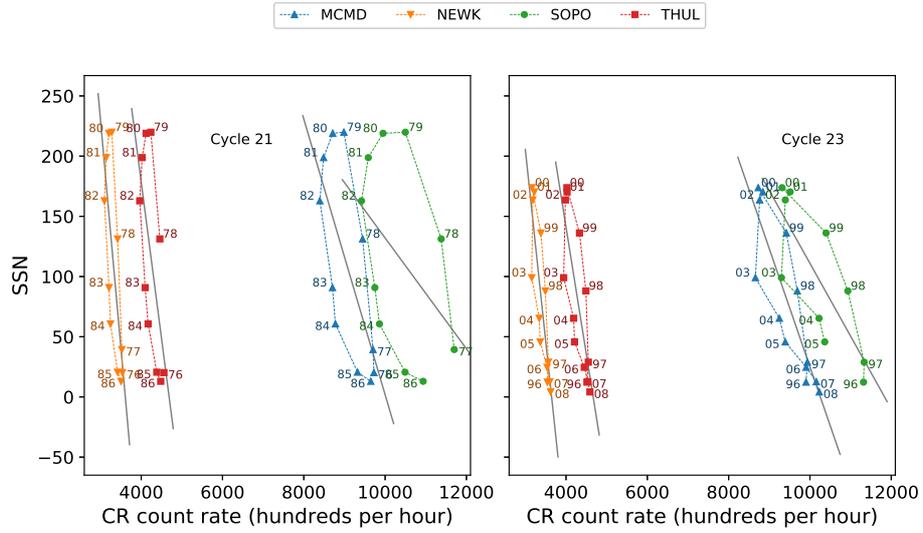}
    \caption{The hysteresis plots for odd solar cycles 21 and 23 and the linear regression fit to the data.}
    \label{fig:linear_odd}
\end{figure}

The linear relations for odd solar cycles are less consistent in their agreement with observed CR intensities, with the correlation during cycle 21 as low as 0.34 for South Pole and as high as 0.91 for Thule. Across both of the odd cycles considered in this study linear regression is not as good a representation  of the data as for even cycles. Figure \ref{fig:linear_odd} shows the wider hysteresis loops which is a characteristic of odd solar cycles and shows visually that a linear fit does not provide a good representation of the data. 

In agreement with the results of \cite{inceoglu_modeling_2014}, it can be seen from the results in Table \ref{table:hysteresis_20-23} that the ellipse models provide estimates of the CR intensity which are in good agreement with the measured intensities due to the increased correlation coefficient for each station during cycle 21 and cycle 23; for SOPO during cycle 21 the increase in $\rho$ is seen to be 0.58 proving the benefit of the ellipse model.

\begin{figure}
	\includegraphics[width=\columnwidth]{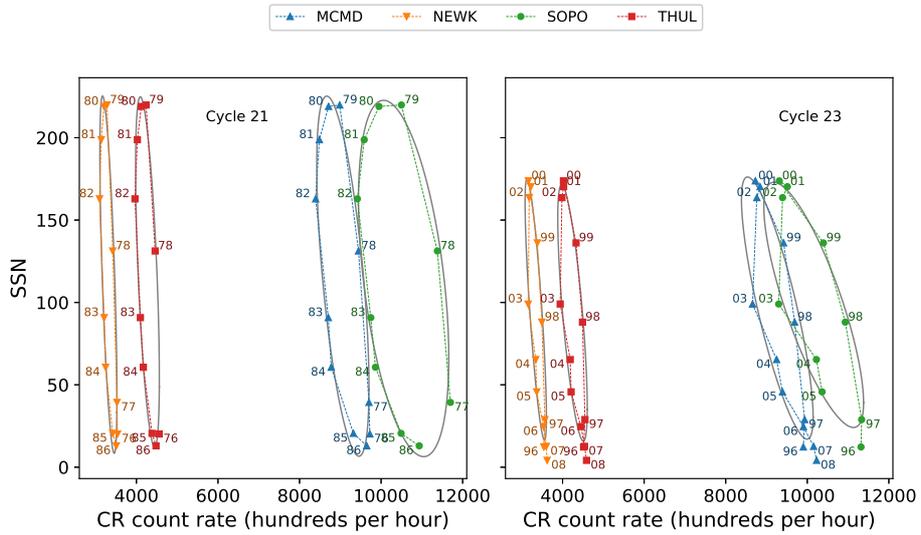}
    \caption{The hysteresis plots for odd solar cycles 21 and 23 and the ellipse fit to the data.}
    \label{fig:ellipse_odd}
\end{figure}

If cycle 24 follows the pattern between odd and even cycles, it is expected that the best fit will be provided by the linear model; however it can be seen in Figure \ref{fig:all_hysteresis} that cycle 24 appears to display a wider hysteresis loop than the two preceding even-numbered cycles. Both the linear model and ellipse model were applied to the hysteresis plots for solar cycle 24 to determine which model would provide the better fit; the correlation between measured CR intensities and modeled CR intensities are presented in table \ref{table:hysteresis_24}.

\begin{table}[!ht]
\caption{Correlation coefficients of the linear regression and ellipse modeling of the hysteresis plots for solar cycle 24.}
\label{table:hysteresis_24}
\begin{tabular}{l c c c c c c c c}
	\hline 
	{} & \multicolumn{2}{c}{Cycle 24} \\
	{} & {Linear} & {Ellipse} \\ \hline
	{McMurdo} & {0.903} & {0.927} \\
	{Newark} & {0.936} & {-} \\
	{South Pole} & {0.883} & {-} \\
	{Thule} & {0.873} & {0.936} \\ \hline
\end{tabular}
\end{table}

The linear model for cycle 24 shows a good correlation between the observed and modeled CR intensities providing evidence to suggest that cycle 24 follows the two preceding even-numbered cycles. The ellipse model does however improve the relation between the observed and modeled CR intensities for 2 out of the 4 stations: McMurdo and Thule. For South Pole and Newark the ellipse model was not able to provide a fit at all, which is believed to be due to the Newark data points crossing where the semi-major axis would be defined for the ellipse model causing the calculation of the semi-major and semi-minor axes to return as not a number, and the South Pole has data missing at the beginning of cycle 24.

\begin{figure}[ht!]
	\includegraphics[width=\columnwidth]{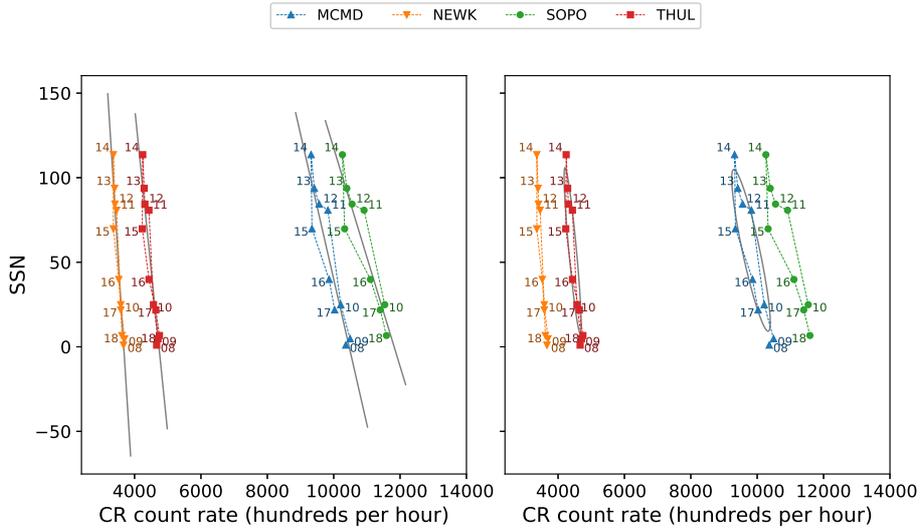}
    \caption{The hysteresis plot for solar cycle 24, and the linear regression fit to the data (left) and ellipse fit to the data (right).}
    \label{fig:24}
\end{figure}

The results for solar cycle 24 do not provide a conclusive answer as to whether cycle 24 behaves like past even solar cycles or odd cycles from this data set alone; however the ellipse model does not provide as significant an improvement for the two modeled NM stations as for odd cycles. The small improvement in the ellipse fit is likely due again to the effects of the extended declining phase of cycle 23 and the unusually low activity of cycle 24.

We repeated the analysis for the additional 4 NM stations that featured in this study. Again, the linear model provided a good fit to the data however the ellipse model was not able to provide a fit; favouring the conclusion that cycle 24 is best represented by a simple linear model as was true for the preceding even-numbered cycles.

Despite cycle 24 having not yet declined to a minimum, it is clear from the observations shown in the hysteresis plots that further data in cycle 24 is unlikely to broaden the loop any further. The hysteresis loop begins to tighten up after 2016 following the broadening between 2014-2016; hence it appears unlikely that by the end of cycle 24 further observations will support the ellipse model and instead will favour the linear fit to the data.

\section{Conclusions}

As cosmic rays are modulated by the heliosphere during the 11-year solar activity cycle, and this effect has been studied for previous solar cycles, the principal aim of this study was to investigate the nature of GCRs during the current activity cycle 24 as it draws to a minimum.

In this study we presented a time-lag analysis between GCR intensity and SSN which showed that cycle 24 has a longer lag (2-4 months) than the preceding even-numbered solar activity cycles (typically 0-1 months); however its lag is not as large as preceding odd-numbered cycles, and cycle 24 follows the trend of a short or near-zero lag for even-numbered cycles. We suggest here that the cause of the extended lag in cycle 24 compared to previous even-numbered cycles is likely due to the deep, extended minimum between cycle 23 and 24, and the low maximum activity of cycle 24 \citep{broomhall_helioseismic_2017}.



It has been previously shown in the literature that there is a striking difference in the shape of the plot of SSN and GCR intensity between odd-numbered and even-numbered solar cycles. Due to the difference in the shape of the hysteresis plots for odd-numbered and even-numbered cycles, we have modeled the hysteresis plots using both a simple linear model and an ellipse model. The results of this study tend to support that cycle 24 follows the same trend as preceding even-numbered cycles and is best represented by a straight line rather than an ellipse, such is the case for odd-numbered activity cycles.

We emphasise that although cycle 24 has not yet ended, the shape of the hysteresis plots suggest that we are now past the main broadening region and the inclusion of further data for cycle 24 will very likely only support the linear model. This study will continue to follow the evolution of the cycle 24 until the onset of cycle 25, in around 2019-2021 \citep{howe_signatures_2018, upton_updated_2018, pesnell_early_2018}, when an update on the final results of cycle 24 should be provided.

\begin{acks}


We acknowledge the NMDB database (\url{www.nmdb.eu}), founded under the European Union's FP7 programme (contract no. 213007) and the teams of the McMurdo, Newark, South Pole, Thule, Climax, Oulu, Kerguelen, and Hermanus neutron monitors for providing data. The neutron monitor data from McMurdo, Newark/Swarthmore, Thule are provided by the University of Delaware Department of Physics and Astronomy and the Bartol Research Institute. The neutron monitor data from South Pole are provided by the University of Wisconsin, River Falls. Kerguelen neutron monitor data were kindly provided by Observatoire de Paris and the French polar institute (IPEV), France. This work uses sunspot data from the World Data Center SILSO, Royal Observatory of Belgium, Brussels. The authors would like to acknowledge the support of the UK Science and Technology Facilities Council (STFC). Funding for the Stellar Astrophysics Centre (SAC) is provided by The Danish National Research Foundation (Grant DNRF106). This research also made use of the open-source Python packages: Numpy \citep{oliphant_guide_2006}, Pandas \citep{mckinney_data_2010},  SciPy \citep{jones_scipy_2001}, and Matplotlib \citep{hunter_matplotlib_2007}.

\end{acks}




\bibliographystyle{spr-mp-sola}
\bibliography{refs_bbt_small}

\IfFileExists{\jobname.bbl}{} {\typeout{}
\typeout{****************************************************}
\typeout{****************************************************}
\typeout{** Please run "bibtex \jobname" to obtain} \typeout{**
the bibliography and then re-run LaTeX} \typeout{** twice to fix
the references !}
\typeout{****************************************************}
\typeout{****************************************************}
\typeout{}}

\end{article} 

\end{document}